\documentstyle[prb,aps,epsfig,draft,floats,amsfonts]{revtex}

\begin{document}

\twocolumn \psfull \draft

\wideabs{

\title{Quest for rare events in mesoscopic disordered metals}
\author{Branislav K. Nikoli\' c}
\address{Department of Physics, Georgetown University,
Washington, DC 20057-0995}

\maketitle

\begin{abstract}
The study reports on the first large statistics numerical
experiment searching for rare eigenstates of anomalously high
amplitudes in three-dimensional diffusive metallic conductors.
Only a small fraction of a huge number of investigated
eigenfunctions generates the far asymptotic tail of their
amplitude distribution function. The relevance of the
relationship between disorder and spectral averaging, as well as
of the quantum transport properties of the investigated mesoscopic
samples, for the numerical exploration of eigenstate statistics is
divulged. The quest provides exact results to serve as a
reference point in understanding the limits of approximations
employed in different analytical predictions, and thereby the
physics (quantum vs semiclassical) behind large deviations from
the universal predictions of random matrix theory.
\end{abstract}

}

\narrowtext

Despite years of extensive studies, initiated  by the seminal work
of Anderson,~\cite{anderson} the problem of quantum particle in a
random potential (``disorder'') still poses new challenges. The
biggest shift in approach and intuition came with the advent of
mesoscopic quantum physics~\cite{mqp}---from the critical
phenomena like description~\cite{gang4} of the
localization-delocalization (LD) transition, to realization that
complete understanding of quantum interference effects,
generating LD (zero-temperature) transition for strong enough
disorder, requires to study the full distribution
functions~\cite{janssen} of relevant physical quantities (such as 
conductance, local density of states, current relaxation times,
etc.~\cite{lerner}) in finite-size disordered electronic systems.
Even in the diffusive metallic conductors, the tails of such
distribution functions show large deviations from the ubiquitous
Gaussian distributions, expected in the limit of infinite
dimensionless zero-temperature conductance $g=G/(e^2/\pi\hbar)$.
These asymptotic tails are putative precursors of the developing
Anderson localization, occurring at~\cite{gang4} $g \sim 1$.

\begin{figure}
\centerline{
\psfig{file=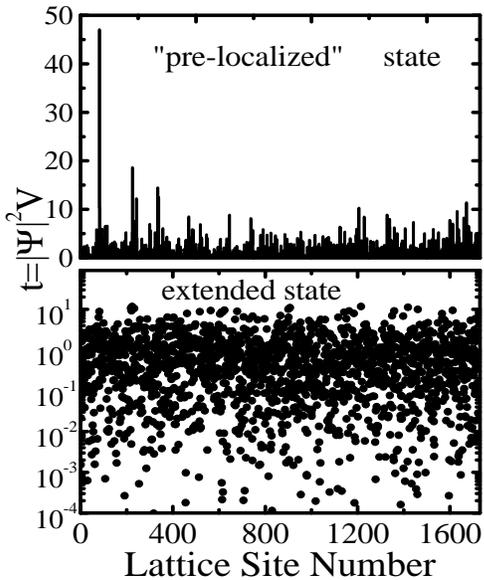,height=3.0in,width=2.5in,angle=0} }
\vspace{0.2in} \caption{An example of eigenstates in the band
center of a delocalized phase in the Anderson model of $W=5$. The
average conductance at half filling is $g \approx 12$, entailing
anomalous rarity of the ``prelocalized'' states. For plotting of
the eigenfunction values in 3D, the sites ${\bf m}$ of $12^3$
lattice are mapped onto $\{1,...,1728\}$ in a lexicographic
order, i.e., ${\bf m}\equiv(m_x,m_y,m_z) \mapsto
144(m_x-1)+12(m_y-1)+m_z$.} \label{fig:state}
\end{figure}

A conjecture about unusual eigenstates, being microscopically responsible
 for the asymptotic tails~\cite{janssen} of
various distribution functions related to transport, was put
forward early in the development of the mesoscopic
program.~\cite{lerner} However, it is only recently~\cite{mirlin}
that eigenstate statistics have come into the focus of mesoscopic
community. Similar investigation in the guise of quantum
chaos~\cite{mqp,ghur} started a decade earlier (leading to the
concept of scarring~\cite{kaplan} in quantum chaotic wave
functions). Thus, the notion of ``prelocalized'' states has
emerged from the studies of quantum disordered
systems~\cite{lerner,muz,efetov} of finite $g$. In
three-dimensional (3D) conductors these states have sharp
amplitude peaks on the top of a homogeneous background. Their
appearance, even in good metals, has been viewed as a sign of an
incipient localization.~\cite{lerner} However, the
non-universality of the prelocalized states tempers this
view,~\cite{smolyarenko} nonetheless, pointing out to the
importance of non-semiclassical effects~\cite{smolyarenko} in
systems whose typical transport properties are properly described
by semiclassical theories [including the perturbative quantum
corrections~\cite{efetovbook} $\sim {\mathcal O}(g^{-1})$]. Here I
provide an insight into the weirdness of such states:
Fig.~\ref{fig:state} plots the amplitude spikes of an anomalously
rare prelocalized state, to be contrasted with the amplitudes of
an ordinary extended state shown in the same Figure. The study of
prelocalized states is not only revealing mathematical
peculiarities of the eigenproblem of random Hamiltonians, but is
relevant for various quantum transport experiments, such as e.g.,
tunneling experiments on quantum dots where coupling to external
leads depends sensitively on the local properties of wave
functions.~\cite{dots} By exploiting the correspondence between
the Schr\" odinger and Maxwell equations in microwave cavities,
it has become possible to probe directly the microscopic
structure of quantum chaotic or disordered wave
functions.~\cite{sridhar}

\begin{figure}
\centerline{ \psfig{file=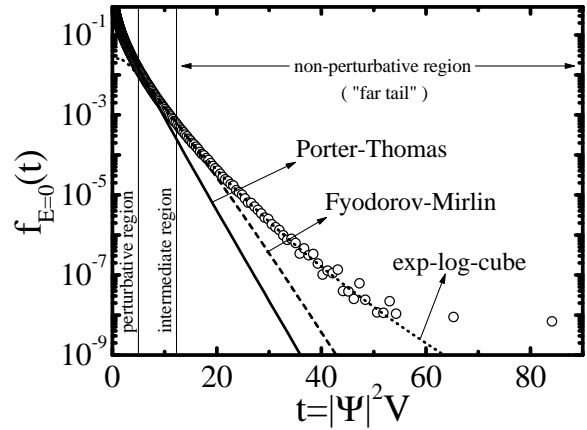,height=3.0in,angle=-90} }
\vspace{0.2in} \caption{Statistics of eigenfunctions in the band
center of the Anderson model with diagonal disorder $W=5$ on a
simple cubic lattice $12^3$. The numerically calculated
distribution $f_{E=0}(t)$ is fitted with: leading order of the
Fyodorov-Mirlin correction, $f_{\rm PT}(t)[1+0.04(3/2-3t+t^2/2)]$,
to PT distribution (valid in the perturbative region); and
exp-log-cube, $0.0256\exp (-0.239 \, \ln^3 t)$, at large
deviations  from $f_{\rm PT}(t)$ in the non-perturbative
region---similar fit is achieved with exp-log-cube,
$0.045\exp[-0.19 \, \ln^3 (t/0.72)]$, akin to $f_{\rm NLSM}(t)$.}
\label{fig:ftini}
\end{figure}

This study presents a numerical result for the statistics of
eigenfunction ``intensities'' $|\Psi_\alpha({\bf r})|^2$ in
closed 3D mesoscopic conductors which are diffusive $L \gg \ell$
($L$ and $\ell$ being the size of the system and elastic mean
free path, respectively), metallic $g \gg 1$, and weakly
disordered $k_F \ell \gg 1$ ($k_F$ is the Fermi wave vector). The
statistical properties of eigenstates are described by a
disorder-averaged distribution function~\cite{efetov}
\begin{equation}\label{eq:ft}
    f_E(t)=\frac{1}{\rho(E) N} \left \langle \sum_{{\bf r},\alpha}
    \delta(t-|\Psi_\alpha({\bf r})|^2 V) \delta(E-E_\alpha)
  \right \rangle,
\end{equation}
on $N$ discrete points ${\bf r}$ inside a sample of volume $V$.
Here $\rho(E) = \langle \sum_\alpha \delta(E-E_\alpha) \rangle$
is the mean level density at energy $E$, and $\langle \ldots
\rangle$ denotes disorder-averaging. Finite-size
disordered samples are modeled by a tight-binding Hamiltonian
\begin{equation}\label{eq:tbh}
  \hat{H}=\sum_{\bf m} \varepsilon_{\bf m}  |{\bf m} \rangle \langle {\bf m}| +
   \sum_{\langle {\bf m},{\bf n} \rangle} t_{{\bf m} {\bf n}}|{\bf m} \rangle \langle {\bf
  n}|,
\end{equation}
with nearest-neighbor hopping $t_{{\bf m} {\bf n}}=1$ (unit of
energy) between $s$-orbitals $\langle {\bf r} | {\bf m} \rangle =
\psi({\bf r}-{\bf m})$ located on sites ${\bf m}$ of a simple
cubic lattice of size $L=12a$ ($a$ being the lattice spacing).
Periodic boundary conditions are chosen in all directions. The
disorder is simulated by taking the potential energy
$\varepsilon_{\bf m}$ to be a uniformly distributed random
variable over the interval  $[-W/2,W/2]$. This is a standard
Anderson model of localization,~\cite{anderson} where weak
disorder is introduced by $W=5$ (the model can also be viewed as
a discretized version of the problem of a single particle in a
continuous random potential).
The eigenproblem of~(\ref{eq:tbh}) is solved exactly by numerical
diagonalization. A small energy window $\Delta E=0.007$ is
positioned around $E=0$, picking up $0--3$ states in each of the 30000 
 conductors with different impurity configurations.
Altogether, 43432 states have been collected for the evaluation
of $f_{E=0}(t)$ as a histogram of intensities at all points
inside the sample ($N=12^3$). This large ensemble makes it
possible to obtain a well-defined far tail (Fig.~\ref{fig:ftini}),
stemming from the amplitudes of prelocalized states (the far tail
has been investigated by numerical simulation only in
low-dimensional systems~\cite{uskiprb,muller} where, strictly speaking,
all states are localized even for weak disorder,~\cite{gang4}
albeit with an exponential localization length in 2D).

In both generic quantum chaotic and quantum disordered systems
eigenstates are characterized solely by their energy (the only
constant of motion) instead of a set of quantum numbers. Their
classical counterparts are nonintegrable---in the entire phase
space for ``hard'' chaos.~\cite{ghur} Quantum chaos implies
semiclassical approximation~\cite{ghur} ($\hbar \rightarrow 0$),
and is less efficient~\cite{sridhar,muller} in localizing chaotic
wave functions through scars than localization by quantum
disorder,~\cite{igorchaos} due to interference effects in the
diffusive motion. Since eigenstates and eigenvalues cannot be
obtained analytically, one usually resorts to some statistical
treatment. The unifying concepts in this pursuit come from
approaches like random matrix theory~\cite{ghur} (RMT) and its
justification (note that matrix elements of Hamiltonians of
disordered solids are spatially dependent and therefore do not
satisfy standard assumptions of RMT) through supersymmetric
nonlinear $\sigma$-model~\cite{efetovbook} (NLSM). RMT gives
universal predictions for the level and eigenstate statistics,
which are applicable in the limit $g \rightarrow \infty$. Its
answer for $f_E(t)$, in time-reversal invariant systems, is given
by the Porter-Thomas~\cite{ghur} (PT) distribution function
\begin{equation}\label{eq:porter}
  f_{\text{PT}}(t)=\frac{1}{\sqrt{2 \pi t}} \exp(-t/2).
\end{equation}
Within RMT, $\Psi_\alpha ({\bf r})$ is a Gaussian random variable, thereby
leading to $\chi^2$ distribution of intensities. The NLSM allows
one to study the deviations from the universality limit 
in the finite-size weakly disordered ($k_F \ell \gg 1$) samples, which
turn out to be determined by the properties of classical
diffusion.~\cite{mirlin} The distribution function
$f_{E=0}(t)$ is contrasted with $f_{\rm PT}(t)$ in
Fig.~\ref{fig:ftini}. For $t \gtrsim 5$, a deviation in the tail
starts to develop, eventually becoming a few orders of magnitude
greater probability than RMT prediction for the appearance of
states with high amplitude splashes. However, the disorder is so
weak that large-$t$ limit tail is shorter, and composed of
smaller values of $f_E(t)$, than some quantum chaotic tails
attributed to scarring.~\cite{kaplan,muller} Small deviations
from PT distribution are taken into account through perturbative
(weak localization) correction~\cite{fyodorov}
\begin{equation}\label{eq:pert}
  f_{\rm FM}(t)=f_{\rm PT}(t)\left[1+\frac{\kappa}{2}
  \left(\frac{3}{2}-3t+\frac{t^2}{2}\right) + {\mathcal O}(g^{-2})
  \right],
\end{equation}
derived by Fyodorov and Mirlin (FM)~\cite{fyodorov} for $\ t \ll
\kappa^{-1/2}$. Here $\kappa$ ($\sim L/g\ell$ in 3D) has the
meaning of a classical (time-integrated) return probability for a
diffusive particle.~\cite{mirlin}
\begin{figure}
\centerline{\psfig{file=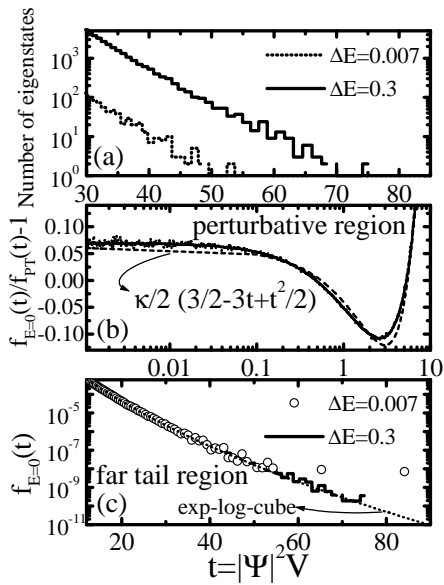,height=3.0in,angle=0} }
\vspace{0.2in} \caption{Comparison of disorder average (over the
impurity ensemble) and combined disorder and spectral averaging
[over small interval $\Delta E$ used to broadened
$\delta(E-E_\alpha)$ from Eq.~(\ref{eq:ft}) into a box function] in
computing $f_E(t)$: (a) histogram of the number of eigenstates
which exhibit the largest amplitude splashes determining the very
end of the far tail of the distribution of $|\Psi_\alpha({\bf
r})|^2$---both histograms are defined by only $1.4\%$ of 43342 or
$1.86 \cdot 10^6$  eigenstates collected in the interval $\Delta
E=0.007$ or $\Delta E=0.3$ around $E=0$, respectively; (b)
perturbative region and FM fit (note a small discrepancy between
the data and the fit, which is increasing upon approaching the
upper boundary $W \simeq 6$ of the semiclassical transport
regime); (c) far tail region and exp-log-cube fit.}
\label{fig:hist}
\end{figure}
In the region of considerable deviations, where $f_E(t)$ decays
much slower than $f_{\rm PT}(t)$, one has to use the
non-perturbative predictions for the asymptotics of $f_E(t)$ in
mesoscopic metals.~\cite{efetovbook} They employ either NLSM
techniques~\cite{mirlin}
\begin{equation}\label{eq:mirlin}
f_{\rm NLSM}(t) \sim \exp \left [- \frac{1}{4 \kappa} \, \ln^3
(\kappa t) \right ],\ t \gtrsim \kappa^{-1},
\end{equation}
or ``direct optimal fluctuation method'' of
Ref.~\onlinecite{smolyarenko} (where only the leading term of
exp-log-cube asymptotics is evaluated
explicitly~\cite{smolyarenko})
\begin{equation}\label{eq:smoly}
f_{\rm DOF}(t) \sim \exp \left [- C_{\rm DOF} \, k_F\ell \,
\ln^3 t \right ],
\end{equation}
aiming to describe the short-scale (and non-semiclassical)
structure of the solutions of Schr\" odinger equation in a
white-noise disorder. The intermediate region of amplitudes
(Fig.~\ref{fig:ftini}) is covered in the NLSM formalism by
\begin{equation}\label{eq:inter}
f(t) \simeq \frac{1}{\sqrt{2 \pi t}}\exp \left[ \frac{1}{2}
\left(-t+\frac{\kappa t^2}{2} + \cdots \right) \right ],\
\frac{1}{\sqrt{\kappa}} \lesssim t \lesssim\frac{1}{\kappa},
\end{equation}
where a correction in the exponent is large compared to unity,
but small compared to the leading RMT term.~\cite{efetov}

Numerical techniques provide the means to solve exactly the
(discretized) Schr\" odinger equation for a quantum particle in an
arbitrary strong random potential, i.e., for all transport
regimes---from semiclassical to strongly
localized.~\cite{nikolic_qc} Such solutions are a good reference
point from which one can gauge analytical predictions and their
``neglect'' of those effects which are hard to deal within a
specific formalism. Analogous numerical investigations of
$f_E(t)$ in quantum chaotic systems have typically dealt with a
single set of eigenstates~\cite{muller} since impurity ensemble
is absent in standard clean examples of quantum chaos (like e.g.,
quantum billiards~\cite{sridhar}). Even in quantum disordered
systems, the computation of $f_E(t)$ has involved only relatively
small ensembles of disordered conductors, or combined disorder
and spectral averages (the disorder-averaging is supposed to
improve the statistics, while claiming that individual eigenstate
would show more or less the same behavior~\cite{muller}). Thus, 
the natural questions arise: Is an enormous ensemble of disordered conductors
indispensable to obtain complete $f_E(t)$ in a good metal? Can spectral 
averaging [i.e., wide broadening of $\delta
(E-E_\alpha)$ in Eq.~(\ref{eq:ft}) into a box function of the
width $\Delta E$] be used to facilitate this quest? Both
questions demand an empirical answer, which is readily obtained
from Fig.~\ref{fig:hist}. Only a small fraction of all
investigated eigenfunctions exhibits the highest observed
amplitudes [panel (a) in Fig.~\ref{fig:hist}]. Therefore, one has
to search for special configurations of disorder where quantum
interference effects can lead to large wave function
inhomogeneities. They are anomalously rare in weakly disordered
conductors, i.e., the tail of $f_E(t)$ is not generated by a
large number of more or less equal-weight impurity
configurations.~\cite{smolyarenko} The effect of spectral
averaging is not {\em a priori} obvious.~\cite{prange} 
Namely, the short scale physics to be discussed below
can correlate eigenstates with neighboring eigenvalues and
therefore produce artifacts of the numerical procedure employed
to get $f_E(t)$ in a finite-size system with discrete spectrum,
when the original definition~(\ref{eq:ft}) in terms of delta
functions is abandoned. Nonetheless, the comparison of $f_E(t)$
from Fig.~\ref{fig:ftini} and the one obtained from the same
impurity ensemble but with $\Delta E=0.3$ (where 62 states are
collected, on average, from each sample) suggests that combined
impurity ensemble and reasonable spectral averaging might be the
most efficient way to obtain a long and smooth far tail
(Fig.~\ref{fig:hist}).

Mesoscopic transport properties of finite-size samples can be
fully delineated  from the ``measurement'' on a computer. This
offers a simple way of comparing the above formulas to the
observation in real systems---instead of trying to deduce the
functional form of $f_E(t)$ phenomenologically (which is usually
inconclusive because of the possibility to fit reasonably-well
different functions), one can use the ``measured'' values, adjust
the free parameters, and, if the analytical expressions work,
explore the viability of physics behind the raw numbers. The
quantitative understanding of transport is also prerequisite for
another reason. The lattice size sets a lower limit on the
disorder strength which ensures the diffusive transport regime
($L \gg \ell$). When disorder is too strong, semiclassical
parameters become nonsensical ($\ell < a$), although the
conductor might still be far away ($g \gg 1$) from the LD
transition point.~\cite{gang4,allen} On the other hand, only for
strong enough disorder the far tail is reasonably long
(Fig.~\ref{fig:ftini}) to allow the extraction of reliable
fitting parameters. The interplay of these three limits leaves a
narrow range $4 \lesssim W \lesssim 6$ of possible disorder
strengths in this study.

The exact zero-temperature disorder-averaged conductance is
calculated from the two-probe Landauer formula $g(E_F)=\text{Tr}
\, [{\bf t}(E_F) {\bf t}^{\dag}(E_F)] \approx 12$ (at
$E_F=0$).~\cite{allen} The Fermi wave vector $k_F \approx 2.8/a$,
averaged over the Fermi surface $E_F=0$ of a simple cubic
lattice, serves as a counterpart of $k_F$ used in theoretical
simplifications which assume a Fermi sphere. For $W=5$ disorder 
$g$ is still dominated by the semiclassical
effects.~\cite{allen} Thus, I use the Bloch-Boltzmann formalism
(applicable when $k_F\ell \gg 1$), in Born approximation for the
scattering on a single impurity, to obtain~\cite{allen} $\ell
\approx 1.4a$ ($k_F\ell \approx 4$).

Small deviations (both in the body and in the tail) of the
numerical $f_{E=0}(t)$ from the PT distribution are well-explained
by $f_{\rm FM}(t)$ [panel (b) in Fig.~\ref{fig:hist}]. However,
$\kappa \approx 0.08$ extracted from the fit is much larger than
the diffusive and universal~\cite{fyodorov} (i.e., independent of
the details of disorder), $\kappa_{\rm diff} = (2/g \pi^2) \,
\sum_{\bf q} \exp(-{\mathcal D}{\bf q}^2\tau)/{\bf
q}^2L^2=0.018$. The exact value of the sum over the diffusion
modes (${\bf q}$ being quantized by periodic boundary conditions)
is evaluated numerically where exponential cutoff provides the
necessary ultraviolet regularization at $|{\bf q}| \sim \ell^{-1}$
(${\mathcal D}=v_F^2\tau/3$ is a classical diffusion constant).
The divergence of the sum in 3D necessitates to go beyond the
diffusion approximation of the NLSM, implying that short-range spatial
correlations become important. This is to be contrasted with the
long-range correlations (described by the soft modes of NLSM) in
low-dimensional systems where $\kappa_{\rm diff}$ is convergent
parameter-free number. Therefore, to reproduce the fitted
$\kappa$, the other relevant (``ballistic'' and non-universal)
contribution $\kappa_{\rm ball} \sim (k_F\ell)^{-1}$, arising
from quantum dynamics on time scales shorter than elastic mean
free time~\cite{mirlin,blanter} $\tau$, has to be included
$\kappa=\kappa_{\rm diff}+\kappa_{\rm ball} \sim (k_F\ell)^{-2} +
(k_F\ell)^{-1}$. The intermediate region is poorly described by
the formula~(\ref{eq:inter}).

The beginning of the far tail $t \gtrsim \kappa^{-1} \simeq 12$
is located approximately using the fitted $\kappa$ from the
perturbative region. The exp-log-cube formulas (\ref{eq:mirlin}),
(\ref{eq:smoly}) imply that this region should be explained by an
exponential function of a cubic polynomial, $C_p \exp (-C_3\ln^3 t
- C_2 \ln^2 t - C_1 \ln t)$. The fit, however, cannot give the
accurate values for all four parameters simultaneously. On the
other hand, the fit including just the cubic term $C_p\exp(-C_3
\ln^3 t)$ works well (Fig.~\ref{fig:ftini},~\ref{fig:hist}),
giving $C_3 = 0.239$ and $C_p = 0.0256$. The NLSM universal result
goes beyond predicting only the prefactor of the leading log-cube
term by including the lower powers of $\ln t$. This form also
fits the data reasonably well with $\kappa=1.39$ and log-cube
prefactor $C_{\rm NLSM} \approx 0.19$ [which is close to the
prefactor $1/4\kappa=0.18$ in $f_{\rm NLSM}(t)$]. Nevertheless,
there are two major puzzles with $f_{\rm NLSM}(t)$ fitting the
numerical data: (1) inasmuch as $\kappa$ is a characteristic
quantity of classical diffusive dynamics, determining both
perturbative and non-perturbative corrections (within the
framework of  NLSM) to RMT picture,~\cite{mirlin} the different
values, required here for different intervals of amplitudes in
Fig.~\ref{fig:ftini}, are hard to reconcile (regardless of the
dynamical origin of $\kappa$); (2) since $\kappa_{\rm diff} \sim
(k_F\ell)^{-2}$, the log-cube prefactor has the form $1/4\kappa
\sim (k_F\ell)^2$. However, the investigation of different
ensembles for a range of $\ell$ ($\propto W^{-2}$) hints toward
the prefactor ($C_3$ or $C_{\rm NLSM}$) being close to linear
function of $k_F\ell$ as is the case with~\cite{drugi} $f_{\rm DOF}(t)$ 
prediction (an unexpected sublinear prefactor behavior is seen 
also in 2D Anderson model~\cite{uskiprb}).

Thus, the exp-log-cube formulas {\em could} account for the far
tail of statistics of eigenstates in the Anderson model. However,
their expected parameters are inadequate, suggesting that short
scale effects below $\ell$, which are to be treated fully quantum
mechanically,~\cite{smolyarenko,blanter} are essential (even in
the perturbative region and for large samples). They lead to
strong dependence of eigenstate statistics in 3D on microscopic
details of a random potential.~\cite{mirlin} It remains to be
seen if the log-cube prefactor can be explained by examining the
short-scale physics (thereby going beyond standard diffusive NLSM
corrections, which are semiclassical in nature and
universal)---Anderson model can hardly satisfy some key
assumptions of the ``direct optimal fluctuation
method'',~\cite{smolyarenko} while ballistic NLSM,~\cite{muz} as
yet, has only reproduced the diffusive NLSM
result.~\cite{smolyarenko}

I am grateful to I.~E.~Smolyarenko for initiation into this
problem and A.~D.~Mirlin for important clarifications. Essential
help along the quest was provided by V.~Z.~Cerovski. Financial
support from ONR grant N00014-99-1-0328 is acknowledged.

\end{document}